\documentclass[11pt,a4paper]{article}
\usepackage[margin=1in]{geometry}
\usepackage{graphicx}
\usepackage{caption}
\usepackage{subcaption}
\usepackage{amsmath,amssymb}
\usepackage{hyperref}
\usepackage{authblk}
\usepackage{xcolor}
\usepackage{lineno}

\title{\textbf{Searches in CMS for New Physics in Final States with Leptons}}

\author[ ]{Anureet Kaur\\
\textit{On behalf of the CMS Collaboration}\\
Panjab University, Chandigarh, India}

\date{}

\begin{document}
\maketitle
%\linenumbers

\begin{center}
\textit{Presented at the 32nd International Symposium on Lepton Photon Interactions at High Energies (LP2025), Madison, Wisconsin, USA, August 25--29, 2025.}
\end{center}

\begin{abstract}
Many new physics models, such as the Sequential Standard Model, Grand Unified Theories, models of extra dimensions, or models like leptoquarks or vector-like leptons, predict heavy mediators at the TeV energy scale. We present recent results of such searches in leptonic final states obtained using data recorded by the CMS experiment during Run-II of the LHC.
\end{abstract}

\section{Introduction}
The Standard Model (SM) of particle physics accurately describes known fundamental interactions, yet several observed phenomena including the nature of dark matter, the matter--antimatter asymmetry, and the hierarchy of mass scales remain unexplained. Many extensions of the SM predict new resonant or non-resonant states that couple to leptons or produce final states rich in electrons, muons, and taus. Leptons provide clean experimental signatures, precise momentum resolution, and reduced QCD backgrounds, making them powerful probes in direct searches for physics beyond the SM.

The Compact Muon Solenoid (CMS) experiment~\cite{CMS2008} at the CERN Large Hadron Collider (LHC) has pursued an extensive search program covering prompt, displaced, and soft-lepton topologies. The results summarized here are based on proton–proton collision data recorded at $\sqrt{s}=13$~TeV during Run-II (2016--2018), corresponding to an integrated luminosity of 138~fb$^{-1}$, and in some cases on additional Run-III data at 13.6~TeV. The following sections highlight five representative analyses covering distinct mass regimes and production mechanisms: soft leptons from compressed supersymmetry (EXO-23-017~\cite{EXO-23-017}), Higgs-portal light scalars (EXO-24-034~\cite{EXO-24-034}), inclusive low-mass $\tau\tau$ resonances using scouting data (EXO-24-012~\cite{EXO-24-012}) scalar leptoquarks produced via muon–quark scattering (EXO-24-005~\cite{EXO-24-005}), and ultra-light axion-like particles in $H\to aa\to4e$ decays (EXO-24-031~\cite{EXO-24-031}).

\section{Soft leptons and compressed electroweak SUSY (EXO-23-017)}
\begin{figure}[ht]
\centering
\includegraphics[width=0.50\textwidth]{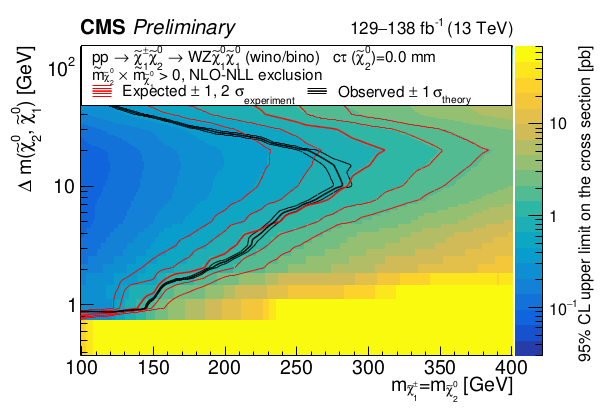}
\hfill
\includegraphics[width=0.45\textwidth]{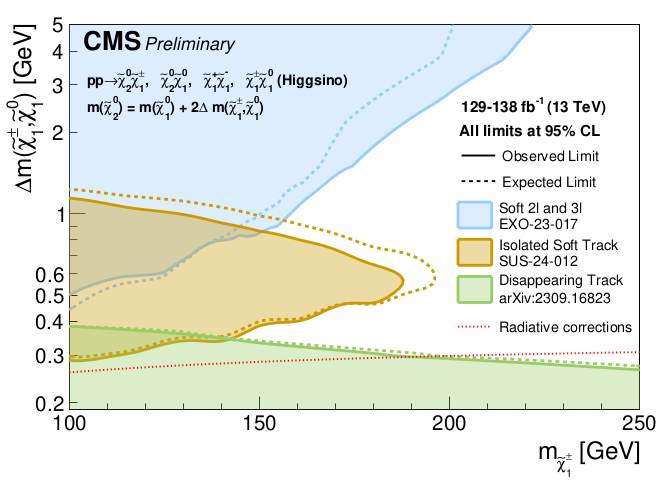}
\caption{
Left: observed and expected 95\%~CL exclusion contours in the $(m_{\tilde{\chi}_1^\pm}, \Delta m(\tilde{\chi}_2^0,\tilde{\chi}_1^0))$ plane for the wino--bino model.
Right: comparison of the observed limit for the higgsino scenario with complementary CMS analyses based on isolated soft tracks and disappearing tracks.}
\label{fig:EXO23017}
\end{figure}

A search is performed for electroweak production of nearly mass-degenerate charginos and neutralinos that yield very soft leptons and missing transverse energy. Dedicated reconstruction extends the reach to electrons with $p_{\mathrm{T}}$ down to about 1~GeV and muons to 3.5~GeV, targeting both prompt and displaced signatures with lifetimes up to 10~cm. The dominant backgrounds, including Drell--Yan, top-quark, and non-prompt sources, are estimated using data-driven template fits. The transverse-mass and dilepton-mass spectra are fitted with empirical functions to search for a local excess. No statistically significant deviation from the background expectation is observed, and limits at the 95\% confidence level (CL) are set on the cross section for simplified wino–bino and higgsino models. For the higgsino scenario, mass splittings $\Delta m(\tilde{\chi}_2^0,\tilde{\chi}_1^0)$ between 0.6~GeV and 50~GeV are probed, excluding higgsino masses up to about 140~GeV, thereby closing the remaining gap left by LEP as shown in Figure~\ref{fig:EXO23017}.

\section{Higgs-portal light scalars in $\mu\mu$ + hadrons (EXO-24-034)}
This analysis investigates an extended Higgs sector in which the SM Higgs boson decays into a pair of new light scalars, $H\to SS\to(\mu^+\mu^-)(h^+h^-)$. Events are reconstructed with a dimuon pair and two hadronic objects consistent with displaced vertices using the full 138~fb$^{-1}$ Run~2 dataset.
\begin{figure}[ht]
\centering
\includegraphics[width=0.47\textwidth]{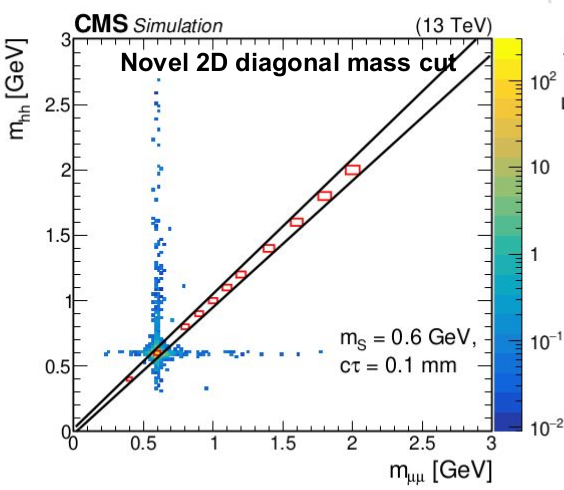}
\hfill
\includegraphics[width=0.47\textwidth]{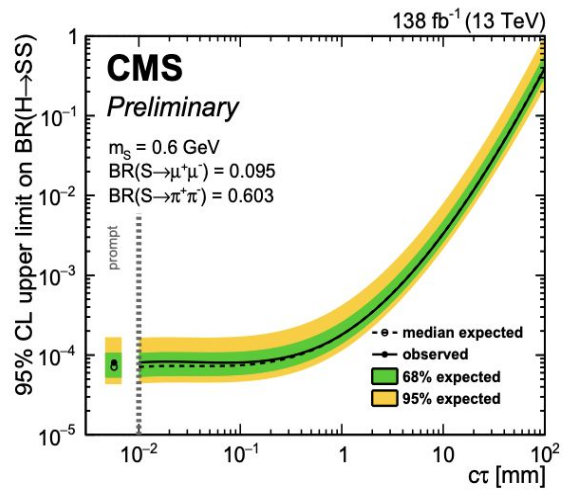}
\caption{Left: two-dimensional distribution showing the diagonal signal region $m_{\mu\mu}\approx m_{hh}$. Right: observed and expected 95\%~CL upper limits on $\mathcal{B}(H\to SS)$ versus scalar mass and lifetime.}
\label{fig:EXO-24-034}
\end{figure}
 A novel two-dimensional diagonal mass requirement, $m_{\mu\mu}\!\approx\!m_{hh}$, suppresses the QCD multijet background by approximately 96\%. The signal is searched for in the four-body invariant-mass distribution across lifetime bins of $c\tau=0.1$--$100$~mm. Backgrounds are estimated from sidebands using an envelope of parametric fits, and systematic uncertainties from trigger and reconstruction efficiencies are included in the likelihood evaluation. No statistically significant deviation from the background expectation is observed, and 95\%~CL upper limits on $\mathcal{B}(H\!\to\!SS)$ are derived, reaching values down to $10^{-5}$ for $c\tau\!\sim\!1$~mm as shown in Figure~\ref{fig:EXO-24-034}. These are the first CMS limits on sub-GeV scalars from Higgs decays within the tracker volume.

\section{Inclusive low-mass $\tau\tau$ with scouting (EXO-24-012)}
\begin{figure}[ht]
\centering
\includegraphics[width=0.48\textwidth]{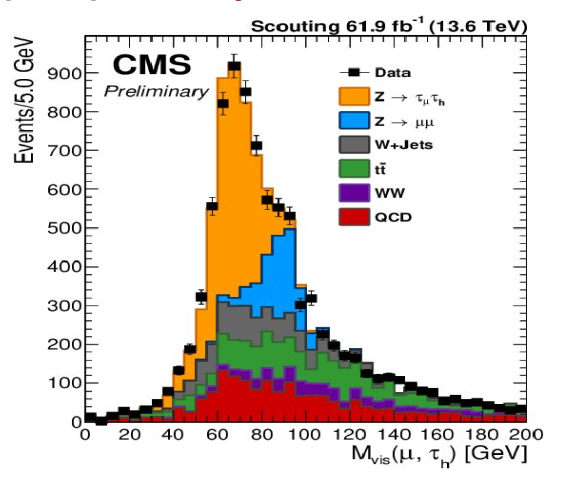}
\hfill
\includegraphics[width=0.47\textwidth]{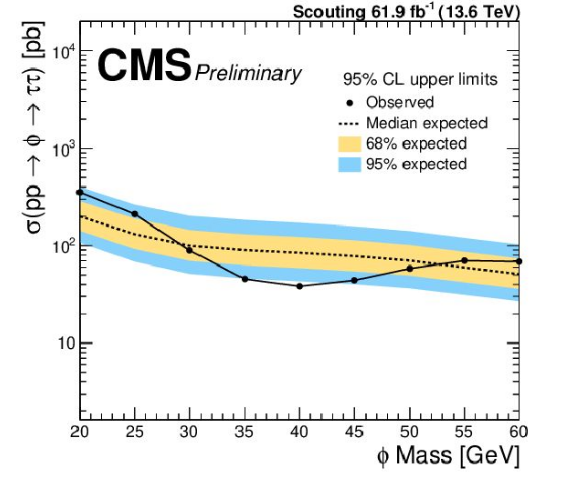}
\caption{Left: distribution of the reconstructed visible mass compared to background prediction. Right: observed and expected 95\%~CL limits on $\sigma(pp\to\phi\to\tau\tau)$ as a function of resonance mass.}
\label{fig:EXO-24-012}
\end{figure}

A search for low-mass resonances decaying to $\tau\tau$ pairs is performed using the CMS scouting dataset recorded in 2022--2023, corresponding to 61.9~fb$^{-1}$ at $\sqrt{s}=13.6$~TeV. The scouting data stream stores reduced event information at high rate, enabling access to the low-mass region (20--60~GeV) not reachable with standard triggers. Hadronic tau candidates are reconstructed down to $p_{\mathrm{T}}\!\approx\!5$~GeV using the Hadron-Plus-Strips algorithm with $\pi^0$ recovery, and a deep neural network (TauNet) provides efficient identification against QCD jets. Events with one muonic and one hadronic tau of opposite charge are selected, and backgrounds from Drell--Yan, $t\bar{t}$, and QCD processes are controlled using data in sidebands. The visible-mass spectrum $m_{\mathrm{vis}}(\tau_\mu,\tau_h)$ is fitted with smooth functions to search for localized excesses. No significant deviation is observed, and 95\%~CL upper limits on $\sigma(pp\to\phi\to\tau\tau)$ reach the order of 10~pb in the 20--60~GeV range as shown in Figure~\ref{fig:EXO-24-012}. This constitutes the first inclusive LHC measurement in this regime, demonstrating the power of scouting data for low-mass resonance searches.

\section{Scalar leptoquarks via $\mu$--quark scattering (EXO-24-005)}
\begin{figure}[ht]
\centering
\includegraphics[width=0.47\textwidth]{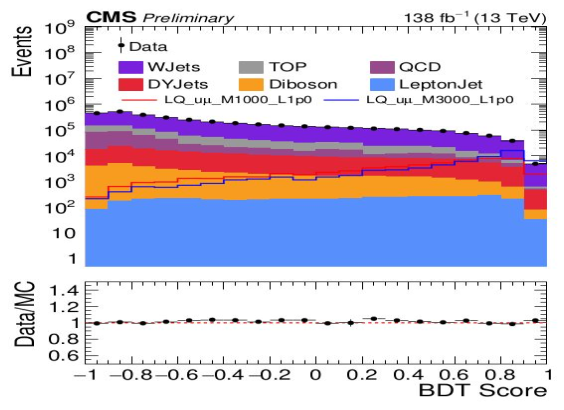}
\hfill
\includegraphics[width=0.50\textwidth]{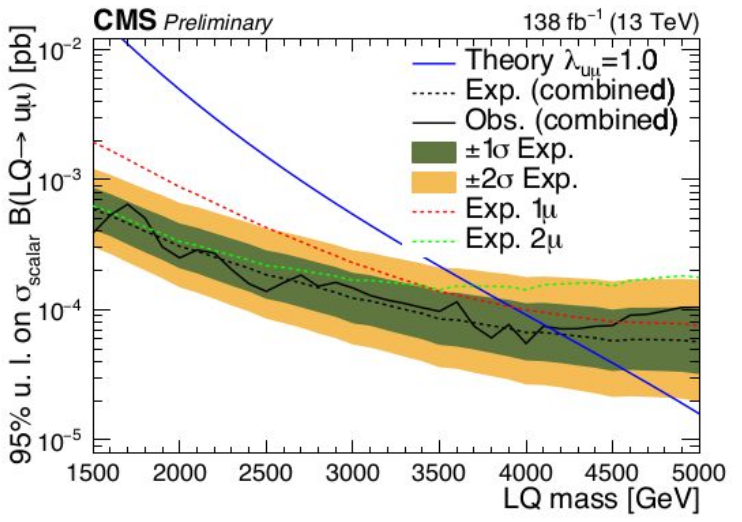}
\caption{Left:distribution of the boosted decision tree (BDT) output score for the combined analysis region, illustrating the agreement between data and the stacked background prediction, with the lower panel showing the Data/MC ratio. 
Right:observed and expected 95\% confidence level upper limits on the production cross section times branching fraction $\sigma(pp\!\to\!{\rm LQ})\,\mathcal{B}({\rm LQ}\!\to\!\mu q)$ as a function of the leptoquark mass, for a coupling value of $\lambda_{u\mu}=1$.}
\label{fig:EXO24005}
\end{figure}

The CMS search for scalar leptoquarks produced via muon–quark scattering explores a unique single-production process predicted in several unification models. The analysis uses 138~fb$^{-1}$ of Run~2 data at $\sqrt{s}=13$~TeV and selects events containing one high-$p_{\mathrm{T}}$ muon and a high-$p_{\mathrm{T}}$ jet, occasionally accompanied by a second softer muon from photon splitting. Signal discrimination is achieved through boosted-decision-tree (BDT) classifiers, and eight final categories are defined according to muon multiplicity, $b$-tag content, and BDT output. The principal backgrounds from $W$+jets, Drell--Yan, and top-quark production are constrained with control samples in data. No excess is seen, and 95\%~CL limits are set on the cross section times branching fraction as a function of leptoquark mass and coupling $\lambda$. Scalar leptoquarks are excluded up to about 5~TeV for large couplings, surpassing previous pair-production constraints and extending CMS sensitivity to new single-production regimes as shown in Figure~\ref{fig:EXO24005}.

\section{Ultra-light ALPs in $H\to aa\to(ee)(ee)$ (EXO-24-031)}

A novel CMS analysis searches for exotic Higgs decays $H\to aa\to(ee)(ee)$, probing axion-like pseudoscalars with masses between 10 and 100~MeV and lifetimes from 1 to 100~$\mu$m. The full 138~fb$^{-1}$ Run~2 dataset is analyzed using a dedicated merged-electron-pair reconstruction technique, in which two Gaussian-sum-filter tracks are matched to a common electromagnetic supercluster to form a merged object. A boosted decision tree is trained to discriminate such pairs from photon conversions and misreconstructed electrons. 

\begin{figure}[ht]
\centering
\includegraphics[width=0.47\textwidth]{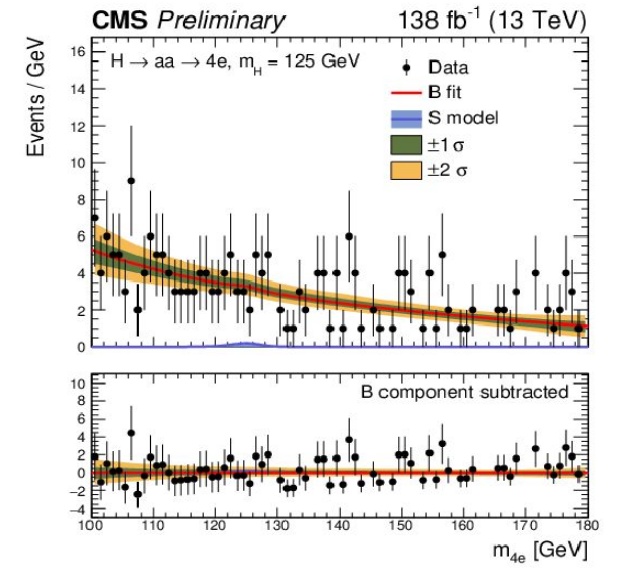}
\hfill
\includegraphics[width=0.48\textwidth]{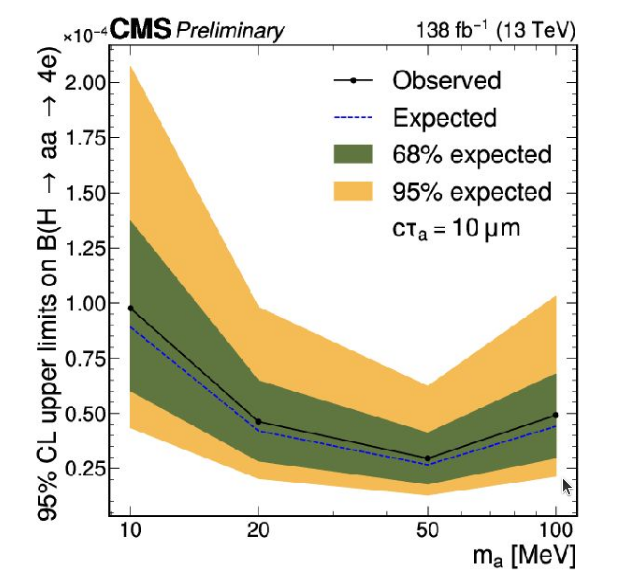}
\caption{Results of the ALP search (EXO-24-031). Left: distribution of the four-electron invariant mass compared with the background-only fit. Right: observed and expected 95\%~CL limits on $\mathcal{B}(H\to aa\to4e)$ versus $m_a$ for representative lifetimes.}
\label{fig:EXO24031}
\end{figure}
The primary backgrounds arise from $ZZ\to4e$, Drell--Yan plus photon conversions, and $\gamma$+jets processes. The four-electron invariant-mass distribution is fitted with analytic functions in several lifetime bins. No statistically significant excess is observed, and upper limits at 95\%~CL on $\mathcal{B}(H\to aa\to4e)$ are set at the level of $10^{-5}$--$10^{-6}$ depending on the ALP mass as shown in Figure~\ref{fig:EXO24031}. This result represents the first CMS sensitivity to electron-coupled ALPs at tens of MeV mass scales using fully reconstructible Higgs decays.

\section{Summary and Outlook}
The CMS experiment continues to expand its reach for new phenomena through advanced reconstruction, machine-learning algorithms, and novel data streams. The results presented here span mass scales from MeV to several TeV, covering complementary models including supersymmetry, extended Higgs sectors, leptoquarks, and axion-like particles. No deviations from the Standard Model are observed, and stringent upper limits are set across all analyses. The upcoming High-Luminosity LHC will provide a tenfold increase in luminosity, improved tracking, and real-time reconstruction at the trigger level, further enhancing sensitivity to soft, displaced, and rare leptonic signatures. Together, these studies establish the foundation for future discoveries in the lepton sector.

\section*{Acknowledgments}
The author acknowledges the CMS Collaboration for providing the opportunity to present this work. 
This talk was supported by the Department of Science and Technology (DST) and the Anusandhan National Research Foundation (ANRF), Government of India.

\end{document}